# Implementation of Machine Learning-based DER Local Control Schemes on Measurement Devices for Counteracting Communication Failures

Rajkumar Palaniappan*, Jan-Niklas Ceschlaw*, Stavros Karagiannopoulos⁺, Christian Rehtanz*
*Institute ie³, TU Dortmund University, Dortmund, Germany
⁺Swissgrid AG, Aarau, Switzerland
rajkumar.palaniappan@tu-dortmund.de

*Abstract*—One of the significant challenges linked with the massive integration of distributed energy resources (DER) in the active distribution grids is the uncertainty it brings along. The grid operation becomes more arduous to avoid voltage or thermal violations. While the Optimal Power Flow (OPF) algorithm is vastly discussed in the literature, little attention has been given to the robustness of such centralised implementation, such as the provision of redundant control solutions during a communication failure. This paper aims to implement a machine learning-based algorithm at each Intelligent Electronic Device (IED) that mimics the centralised OPF used during communication failures using IEC 61850 data models. Under normal circumstances, the IEDs communicate for centralised OPF. In addition, the system is trained offline for all operational conditions and the individual look-up tables linking the actual voltages to the DER setpoints are sent to the respective controllers. The regression models allow for the local reconstruction of the DER setpoints, emulating the overall OPF, in case of a communication failure. In addition to the regression control, the paper also explains an offline learning approach for periodic re-training of the regression models. The implementation is experimentally verified using a Hardware-in-the-loop test setup. The tests showed promising results compared to conventional control strategies during communication failures. When properly trained and coordinated, such an intuitive local control approach for each DER could be very beneficial for the bulk power system. This machine learning-based approach could also replace the existing Q(V) control strategies, to better support the bulk power system.

*Index Terms*—Active Distribution Grids, Machine Learning, Data-driven control design, Optimal Power Flow, Communication redundancy

## I. Introduction

All over the world, the energy supply is undergoing significant upheaval where passive grids are turning to active distribution grids (ADG). The share of renewable energy sources (RES) in the gross electricity consumption in Germany is planned to increase up to 65% by 2030. This could potentially increase due to the requirements of the Renewable Energy Sources Act 2021 (EEG) [1], [2]. The penetration of distributed energy resources (DERs), RES in particular, with high volatility, poses various challenges for the distribution system operators (DSO). Such unpredictability leads to a considerable challenge for the DSOs to deal with congestion management [3]. Conventionally, there are limited to no field measurements available in real-time. This is expected to change in the near future, where the DSOs could choose to install more field measurements, thereby better grid control.

Information and communication systems are becoming very prolific in the ADGs. There are several schemes on which this information exchange be facilitated to operate the grid without voltage or thermal violations. While the centralised and distributed control schemes deliver better results for the overall, or part of the grid [4], [5], they are heavily reliant on the communication infrastructure [6]. The local control schemes are self-reliant to decide on the DER response [7]-[9]. While both strategies have their advantages and limitations, data-driven schemes provide the option of designing a control strategy, combining the advantages of both local and centralised control.

In the literature, several publications deal with the data-based control of DERs. In [10], local controls for the DERs, battery storage systems and controllable loads are designed based on the results of an offline and centrally performed Chance-Constrained optimal power flow (OPF). Machine learning techniques such as support vector machines and piecewise linear regression are applied to the obtained dataset. However, this implementation considers a system without installing communication and measurement technology. In [11], further state-of-the-art methods for the data-based design of local control systems are combined. In a case study, two open-loop and two closed-loop methods are compared.

In [12] and [13], the local reconstruction of a central OPF is in focus. In [12], only quantities that guarantee a feedback-free control of the system are used to reconstruct the OPF. The possibility of performing the reconstruction of the OPF at a node based on locally available quantities and allowing limited communication to other nodes is discussed. In [13], multiple linear models are used. The model also includes an optimised choice of predictors composed of variables locally available at a node.

Furthermore, kernel-based non-linear machine learning models are also proposed. In [14], these are used to determine the optimal reactive power ($Q$) infeed of the converters. An overview of the possible machine learning techniques applied for OPF is mentioned in [15].

While there are several solutions available in the market and research for distribution grid automation, the current research

This paper is based upon work in the project i-Autonomous (No. 03EI6001A), within the future-proof power grids initiative supported by the German Federal Ministry for Economic Affairs and Energy (BMWi). (Corresponding author: Rajkumar Palaniappan) focuses on creating a modular system architecture for implementing protection and smart grid control algorithms on a measurement device to be distributed across the grid [16]. The idea is to make the software and the hardware independent of each other, enabling easy portability of the software. The IEC 61850 data models [17], [18] are used for configuring the algorithms and providing a modular configurable application, independent of the hardware. This means that the software can also be ported and can be device-independent. In this implementation, a standardised measurement acquisition system (EPPE CX) is chosen as the hardware [19]. In addition to recording the measurement values, a network of these devices in the distribution grid will perform local monitoring and adapted control of the grid. Smart grid control functions such as State Estimation (SE), OPF and voltage control exist on the devices [20], [21]. For example, it is possible to intervene in the grid by controlling the flexibilities in real-time, based on the grid conditions.

Although the original idea of the research is to deal with controller-field device centralised control, additional and alternative solutions are to be provided for communication redundancy. This setup is done so that all the decentralised devices are responsible for measurements and receiving DER setpoints from the controller. In contrast, the central controller IED executes the SE, OPF algorithms and sends out the DER setpoints. In case of a communication failure, the central OPF might not be available and the field IEDs cannot perform any operation. This could potentially lead to a grid violation. For this purpose, this paper proposes to use offline-simulated OPF data to pre-program the field IEDs and use these datasets when there is a communication failure. The regression tables store information of the central OPF. During a communication failure, based on the actual system state, the field IEDs use these datasets to drive the DER to a setpoint resembling the central OPF. The datasets are updated at regular intervals, such that the new OPF system states are also updated in each IED. The regression algorithm is also compared to a conventional $Q(V)$ control.

The main contributions of this paper are as follows:
1. Implementation of machine learning algorithms on a measurement IED using IEC 61850 standards to create regression tables representing the central OPF.
2. Quantification of the benefits using the proposed method against conventional $Q(V)$ control via a case study.
3. Periodical retraining of the regression look-up tables compared to static values.

The general overview of the proposed methodology is indicated in Fig. 1. As the first step, offline calculations are made. The training data used to model the local regression control are created using the OPF of the MATPOWER extension [22]. The results of the offline simulation are the optimised $P, Q$ setpoints for the DERs, from 0% to 100% in 5% steps (21 steps in total). The next step is the clustering of the training data and the design of the regression algorithm. The machine learning algorithm processes the training data to create 42 datasets (21 for $P$ and 21 for $Q$). These datasets are preloaded to the field IEDs before the simulation start. The final step is the evaluation, where the validity of the regression algorithm is experimentally verified. The test setup includes decentralised IEDs as controller and field IEDs. These IEDs include the control functions, one of them being the centralised OPF, where the controller IED receives measurements from all field IEDs and sends back $P, Q$ setpoints in return. The regression algorithm is tested on a Cigré benchmark grid [23], simulated on the real-time simulator (RTS). The simulation is started with a centralised setup and communication loss is triggered on the RTS. As soon as the field IEDs receive no specific instructions from the controller IED, they use the regression algorithm to decide on the RES setpoints, thereby mimicking the centralised OPF.

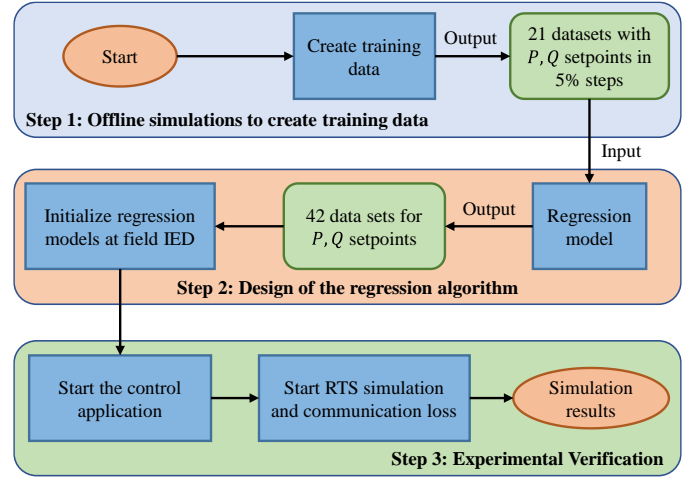

Fig. 1: Overview of the regression algorithm workflow.

The remainder of the paper is organised as follows. Section II gives an overview of the mathematical formulation used in the centralised OPF. In Section III, the methods used to design regression models are explained. Section IV explains the simulation setup for the experimental verification. Section V discusses the results of the RTS simulations and the paper ends with a conclusion and an outlook in Section VI.

## II. CENTRALISED OFFLINE OPF

The standard implementation of a centralised OPF algorithm is used to plan the deployment of controllable RESs. For this purpose, optimisations for the expected grid conditions are carried out based on predictions of the grid loads. These predictions are based on historical grid data, supplemented by additional information, such as weather data. In addition to using the available DER capacity optimally, on-load tap changer stepping can also be used to optimise the operating point.

As explained in the previous section, the idea of the project i-Automate is to collect decentralised measurements. The field IEDs send the real-time measurements to the controller IED, typically situated in the secondary substation. Based on these measurements, the controller IED performs the SE algorithm based on [24],[25] to determine the states of the unmeasured nodes. Using this system state information, a centralised OPF is performed. This capturing of the various possible $P, Q$ setpoints for the RESs is done to train the field IEDs to act independently in case of a communication failure with the controller.

### A. OPF Formulation

The grid information is read at the start of the simulation. It contains the topology information, the DER parameters, the generator costs and the grid configuration. Using this information, an iterative execution of OPF is performed. To include the results of the OPF in an extended operating range, scaling factors are introduced for the load and the available maximum generation power $P_\text{max}$ of the RESs. The DERs of the same type are all scaled by the same factor to limit the calculation complexity. Thus, the correlation of the solar irradiation and the wind speed is assumed perfect so that the same scaling factors are used at each node. Since this is only a benchmark grid, the scaling was considered to be simple. In more complicated grids, the scaling could be extended since typical distribution grids are not more than a few km away and the factors considered here should not be deviant for all DERs. This simplification is made as it is assumed that the DERs are located close to each other.

The three parameters are changed individually and independently during runtime with the scaling factors for the loads and DERs. Since each scaling factor ranges from 0 % to 100 % in steps of 5 % and thus can assume 21 different values, $(21)^3$ different combinations are created in total. Accordingly, the OPF is executed 9261 times. For the OPF, for a grid with $n$ nodes and the set $\sigma$ of generator nodes, an objective function is defined according to equations (1) and (2):

$$C = c_v \sum_{i=1}^{n} f_v(V_i) + c_q \sum_{i \in \sigma}^{n} (Q_i)^2 + c_p \sum_{i \in \sigma, i \neq 1}^{n} (P_{max,i} - P_i)^2 \quad (1)$$

where

$$f_v(V_i) = \begin{cases} 0, & |V_i| \leq 0.01 \\ (|V_i - 1| - 0.01)^3, & otherwise \end{cases} \quad (2)$$

The optimisation function represents the intended behaviour of the OPF. The nodal voltages $V_i$ should be as close as possible to the nominal voltage. A dead-band is introduced to have no cost when the voltages are near 1.0 p.u. and no voltage control is required. The dead-band range could be pre-set based on requirements. The MATPOWER Interior Point Solver (MIPS) is used to solve the OPF problem and it involves the use of the first and second derivatives of the cost function. Because $f_v(V_i)$ is defined in two separate sections, it is vital that the first and second derivatives of the cost functions should be continuous. Furthermore, the $P_i$ infeed of every DER should be close to its presently available $P$ to utilise the DER fully and minimise the required power drawn from the overlying grid. Another quadratic contribution to the cost function is the $Q_i$ injection.

A weighting of the individual partial costs is possible with the help of the cost factors $c_v$, $c_p$ and $c_q$. Since the costs of $P$ curtailment are much more than that of $Q$ control in the distribution grids, it is assumed that the value of $c_q \ll c_p$. The training data consist of the setpoints determined by the OPF for the $P$ injection $P_{\text{SP},i}$ and the $Q$ injection $Q_{\text{SP},i}$ of the currently considered DER as well as the associated nodal voltage $V_i$ and the span angle $\varphi_i$. The setpoints of the powers are calculated in % of the maximum rated $S_\text{max}$.

### B. Boundary Conditions and Cost Functions

The slack grid at node one is modelled as a generator with a high infeed capacity and an equally large reverse power flow capacity to deliver and absorb power. The values are chosen so large that there is practically no limitation. When performing the OPF, the boundary conditions apply to the RESs according to TABLE 1. The value of the scaling factor for the maximum DER power $P_\text{max}$ is represented by the variable $s$. The solution space of the OPF is thus bounded by the currently available $P'_\text{max}$, the power factor and the maximum apparent power $S_\text{max}$. For the non-controllable DERs, it is also specified that the $P$ infeed must correspond to the maximum available $P$. Thus, the operating point of these DERs is fixed and cannot be used by the OPF for further optimisation.

TABLE 1: Boundary conditions for RESs

|  | Controllable RES | Non-controllable RES |
|---|---|---|
| **Power Factor** | $\cos(\varphi) \geq 0.9$ | Fixed power factor $\cos(\varphi) = 1$ |
| **Max. available $P_{max}$** | $P'_{max} = s \cdot P_{max}$ | $P'_{max} = s \cdot P_{max}$ |
| **Min. P infeed** | $P_{min} = 0$ | $P_{min} = P'_{max} = s \cdot P_{max}$ |

The parameters are configured accordingly to generate the training data for the Cigré grid model. The parameters chosen for the parameterisation of the boundary conditions and cost function of the OPF are as follows:

- Power factor $\cos(\varphi) = 0.9$
- Cost parameter for the voltage term $c_v = 2\text{e}^3$
- Cost parameter for the reactive power term $c_q = 1\text{e}^4$
- Cost parameter for the active power term $c_p = 1\text{e}^6$

The parameter $c_q$ is selected to be lower than $c_p$ by a factor of 100. The coefficients are selected to achieve a good trade-off between all the objective terms. The individual terms of the cost function must be weighted such that the optimisation objective is represented. In this case, the costs for $P$ curtailment is higher than the cost for $Q$ control of the DERs.

### III. DESIGN OF THE REGRESSION MODEL

While any particular artificial intelligence algorithm could be applied to the hardware prototype, three specific machine learning algorithms were selected for implementation. The training data created will be used to design a local control for the DERs. For this purpose, the regression algorithms, which reproduce the training data patterns, are implemented. Before transferring the regression algorithms to the EPPE CX devices, a standalone program is developed to create the regression models for initialising the control.

### A. Learning Algorithms

Three regression algorithms are selected, which differ in their complexity and configuration. The most straightforward approach is a linear regression with a straight line. The data sets are created such that $P, V$ and $\delta$ are available as variables. The next approach was the piecewise linear regression that fixed the nodal voltage as the only variable due to the more complex nature of the calculations. However, it can still be configured and the user can choose the number of breakpoints to be used. For this application, a pattern similar to conventional $Q(V)$ curves in the training data can be assumed for the $Q$ as the system response. Accordingly, the maximum number of breakpoints was limited to two, minimising the computational effort. The regression algorithm then determines the optimal position of the breakpoints. Finally, a fully automated algorithm is available that compares the two aforementioned linear methods and selects the approach that approximates better to the behaviour of the OPF algorithm. Thus, both the optimal number of breakpoints and their optimal position are determined.

The last approach was the *k*-Nearest-Neighbor-Regression, which uses the node span as a variable and generates a look-up table, which defines the course of the regression curve with the data pairs stored there. The number of nearest neighbours $k$ can be configured freely. Nearest-neighbour regression predicts the system response for a given input vector $X$ by averaging a selection of known system responses from the training data.

In the following approach, let m = 1, meaning that the output vector $Y$ contains only one element. From the totality of training data sets $(X_i, Y_i)$, a number $k$ is selected whose $X_i$ is closest to the input $X$. This selection forms the so-called neighbourhood $N_k(X)$. The Euclidean distance between $X_i$ and $X$ can be used to determine $N_k(X)$. The predicted system response $\hat{Y}$ is obtained from the mean of the system responses $Y_i$ of the set $N_k(X)$. This is shown in equation (3), as explained in [26]:

$$\hat{Y}(X) = \frac{1}{k} \sum_{X_i \in N_k(X)}^{n} Y_i \qquad (3)$$

For the selected Cigré benchmark MV model, the twenty closest data points are considered ($k = 20$) and the range of the input data is divided into one hundred sections. With the help of this configuration, a course of the regression curve can be achieved that is as monotonous as possible and without larger jumps. The training data are now plotted here as a function of nodal voltage. For the $Q$ curves, a similarity to the characteristics of the conventional $Q(V)$ curves can be seen in each case. According to the cost function used in equation (1), the $P$ infeed of the DER is level with the currently available $P_{\max}$.

When comparing the regression models, it was noticeable that the nearest-neighbour regression depicts the local properties of the system response more accurately. This can be explained by the fact that the slope of the curves varies locally. In contrast, the straight lines of the (piecewise) linear regressions are influenced by a larger data range and therefore follow a global trend. For this reason, the nearest-neighbour regression was chosen for further analysis in this work. For each operating point, the dependence of the $P, Q$ setpoints on the nodal voltage is modelled, resulting in a total of 42 different regression models for each DER. The resulting $P, Q$ regression models for the wind farm at node 6 based on equations (1) and (2), with both $P_{\max} = 60\%$ and $P_{\max} = 100\%$ is depicted in Fig. 2.

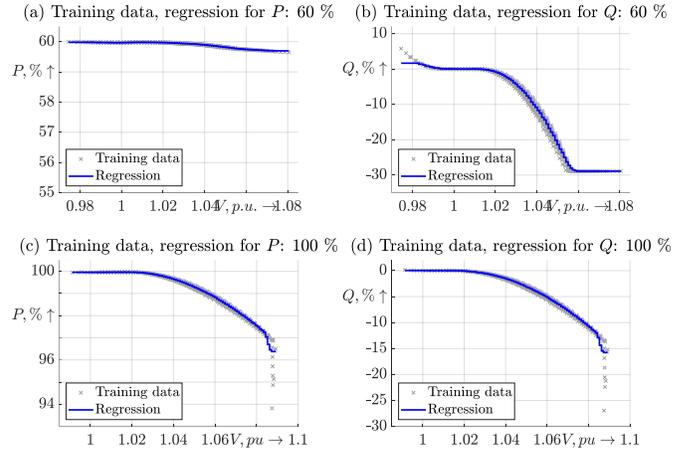

Fig. 2: 20-nearest neighbour regression model with training data for 60% $P_{max}$ and 100% $P_{max}$ setpoints.

### B. Implementation on the EPPE CX devices

The periodic execution of the regression algorithms and the control using the learned regression models are integrated into the existing EPPE CX device program. The implemented functionalities are outlined in Fig. 3. The task area, which includes processing the received setpoints, detecting a communication failure and the subsequent transition to a local control using regression models, is implemented in the client classes.

The training scenarios may not realistically consider all the possible operational conditions. Correspondingly, the actual grid scenario need not always fall under the offline trained grid scenarios. For this reason, the periodic retraining of the regression models is essential. This retraining considers the new grid conditions which were not previously considered during the initial training process. During normal operation, the $P, Q$ setpoints sent by the controller IED are passed on by the field IED and included in the training data, if not previously available.

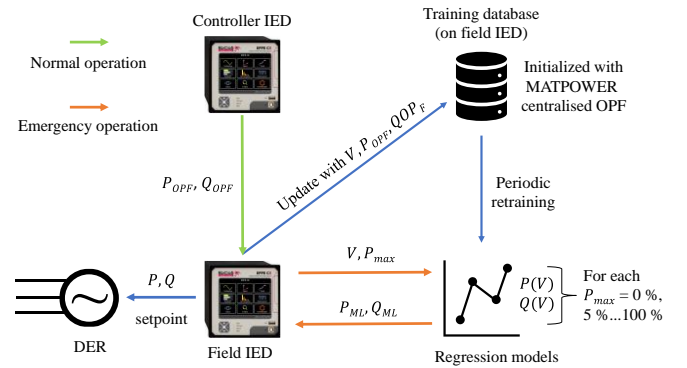

Fig. 3: Processing of setpoints under various operations

In addition to the $P, Q$ setpoints, the last recorded nodal voltage is also written to the data set to be used later as a variable. The measured nodal voltage and the current scaling value are necessary during a communication failure. The scaling value is used to determine the operating point and thus the appropriate regression model to be applied. The nodal voltage is then used as the input value for the regression models, each allowing the learned value to be determined.

The regression models are imported into the IEDs during program start-up. The program recognises the data format of each regression model based on the included identifier. Furthermore, the configuration file contains relevant logical nodes in the data model implementation. When the program is started, the field IEDs are synchronised with the controller IED, enabling the synchronisation of all the field IEDs among themselves. A regular resynchronisation between the IEDs is necessary. There may also be an interruption of the program flow during runtime, which is only resumed after successful resynchronisation. This is always the case if a scheduled resynchronisation after a certain period of time or if a new connection with the controller IED is detected after a connection loss.

All valid setpoints $P_{SP}$ and $Q_{SP}$ received from the central OPF are included in the training data set, where the corresponding training data set is selected based on the actual scaling value. If the respective training data set already contains the maximum number of data points, one or more data points are deleted to fit into the maximum number of entries. The entries to be deleted are determined from those whose nodal voltage is closest to the voltage of the new setpoint. From these entries with $P_i$ and $Q_i$, the one with the most significant value of the criterion formulated in equation (4) is deleted. In the case of a system behaviour change, this procedure can favour adjusting the regression models as old training data are exchanged with new entries.

$$\max \left((P_{SP} - P_i)^2 + (Q_{SP} - Q_i)^2\right) \qquad (4)$$

The training data are only available in a limited range of the theoretically possible nodal voltages. However, an overvoltage or under-voltage situation that is not considered in the training data must be appropriately responded to. For this reason, a local violation of the voltage band is to be checked. If there is no violation, the learned $P, Q$ setpoints are determined using the previously selected regression model and the actual nodal voltage as variables. The implementation uses the derived setpoints and incorporates specific emergency measures for critical system states, for example, $P$ reduction by 10 %.

For the look-up table of an NNR, the closest entry is selected. Suppose an under-voltage is detected ($V < 0.9$ p.u.), then the provision of $P$ and capacitive $Q$ by the RES is set to the available maximum values. If an overvoltage is detected in the grid ($V > 1.1$ p.u.), the setpoints result from the regression models, but the determined operating point of the plant is decremented by a step of 10 % beforehand. Here, a switchover to a regression model of a lower operating point takes place. Suppose the nodal voltage is not below the limit value of 1.1 p.u. again by the next call of the function, the operating point is decremented by another 10 %. This process can be continued until no more $P$ is fed in ($P_{max} = 0\%$). The change to the newly determined setpoints is done step by step as a linear transition to avoid a sudden jump. The upper and lower limits of $Q$ depend on the $P$ setpoint. Either the boundary condition of the minimum allowable power factor or the maximum $S_{max}$ is applied, depending on the condition prescribing the tighter limits. It follows the limitations as in equation (5):

$$P^2 + Q^2 \leq S_{max}^2 \qquad (5)$$

The regression control is explained in Fig. 4.

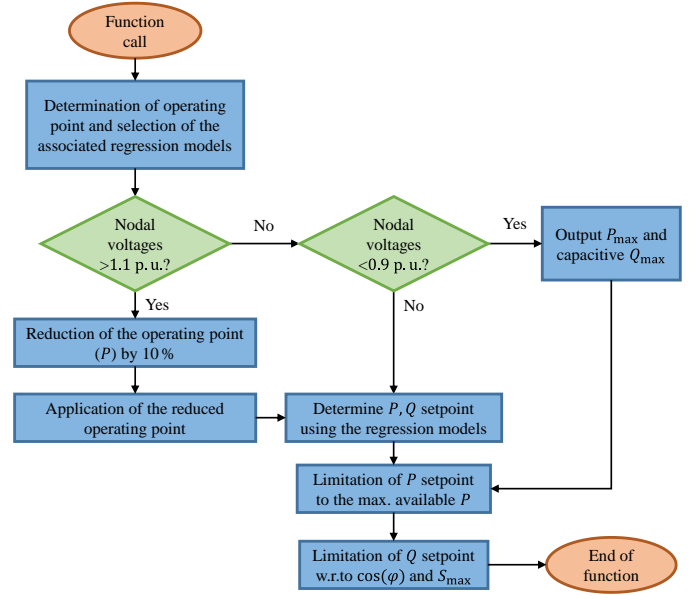

Fig. 4: Regression control sequence

IV. SIMULATION SETUP

Fig. 5 shows the grid model considered, based on the Cigré benchmark MV grid. A few adaptations are made to the grid to fit the requirements, e.g., only two DERs are assumed to be controllable. Measurements are available from the overall grid to maintain complete observability. The RTS simulates the grid and sends the measurements to the controller IED at the primary substation and two field IEDs. These three IEDs are physical hardware IEDs, whereas the other measurement points are virtually represented. The field IEDs also transfer the $P, Q$ setpoints of the RES, which are sent from the controller IED.

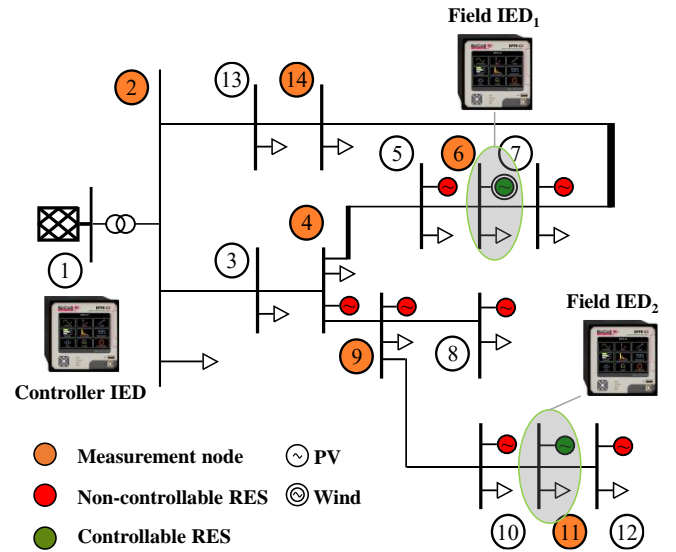

Fig. 5: Test grid based on Cigré Benchmark MV Grid

To ensure that the simulation is dynamic, the grid loads and the available DERs are scaled on the RTS with the help of standard load and generation profiles. The corresponding profile is applied to all loads and all RESs. In this way, the implementation of the profiles is consistent with using the scaling factors mentioned previously. The present values of the PV, wind profile and the measurements are communicated to the IEDs. The RTS sends a signal for synchronisation to the controller when the simulation is started. A MATLAB function is included in the Simulink model, which evaluates the same cost function optimised by the OPF at each simulation time. The comparison of the calculated costs determines the performance of different control strategies. Fig. 6 shows the simulation and communication setup used for the experimental verification.

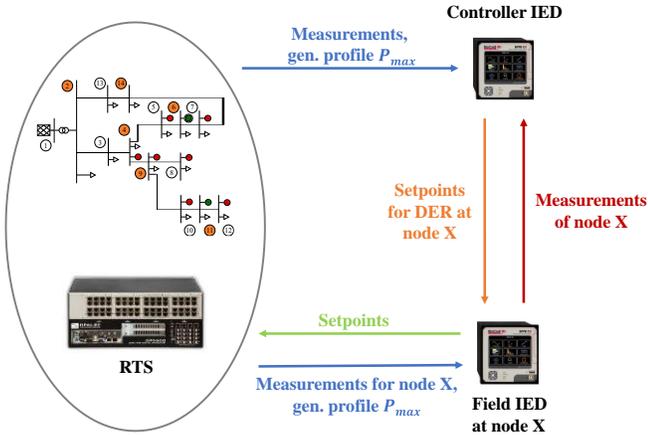

Fig. 6: Simulation and communication setup [27], [28]

The SE and the OPF use standardised logical nodes according to IEC 61850-7-4 [18] for data management. The measurement data is stored, for example, in a logical node with the designation MMXU. For each EPPE CX device, two IEDs must be created in the data model, one of which exists only virtually in the form of the RTS and forwards the measurements to the actual IED. The measurements of all other measured nodes, whose IEDs exist only virtually, are sent directly to the controller IED by the RTS. The logical node of the type MMXU is used to manage the measured data.

Since only two DERs are assumed to be controllable in the grid, the remaining DERs are uncontrolled. It is established by setting up the DERs with a power factor of $\cos(\varphi) = 1$ and means that the $P$ infeed always corresponds to the presently available $P_{\max}$. For the controllable DERs at node 6 and node 11, the $P, Q$ setpoints received from the field IEDs are used instead. These setpoints are either calculated by the controller IED as part of centralised OPF or by the field IEDs using the regression algorithm. The $P$ setpoints are limited at any time to the maximum values specified by the actual scaling value. In comparison, the controllable DERs can also be controlled with a fixed power factor or conventional $Q(V)$ curves during communication failures. To make it comparable to the regression control, the $P$ setpoint is rounded down to the nearest 5 % step.

## V. EXPERIMENTAL VERIFICATION

The experimental verification is divided into two steps. First, the control behaviour of the local regression (in case of communication failures) is examined. The behaviour of the regression control is to be evaluated first to obtain a basic assessment of the performance of the developed approach. This procedure is also called batch learning and can include regular retraining of the regression model as soon as the training data has been updated with a set (batch) of new data sets [29]. Depending on the use case and relevant time constants, suitable specifications must be made for the size of the batches and the execution interval. This also raises the question of how the need for an update can be detected.

As the second step, the effects of the regression model update are examined. For this purpose, the regression models are retrained for a changed system behaviour. Additionally, it is routinely checked if the connection to the controller IED is lost. If the controller IED does not send any new setpoints, the timestamps are not updated. If the timestamp is younger than one minute, the setpoints are valid and forwarded to the DERs. On the other hand, if the timestamp is older than one minute, the communication is assumed to be broken. In this case, the transition to local control takes place, which uses the learned regression models. The program requires approximately 2 seconds to complete the OPF calculations for the selected benchmark grid. Thereby, it can be assumed that OPF will not require more than a few seconds, even for more complicated grid topologies.

The profiles used for the following simulations are based on the time series of the German power grid [30]. The profiles are adjusted for the selected grid model to simulate voltage violations. A $10-day$ period is chosen to verify the learned regression models. Since it is not practical to simulate over ten days, the simulation time is streamlined and determined that the $15-minute$ steps in the time series correspond to $30\ seconds$ simulation on the RTS and the simulation is reduced to eight hours. Some of the essential features of the simulation are given here:
- Controller IED:
  - SE execution cycle: 10 seconds
  - Execution cycle of the OPF: 30 seconds
- Field IED:
  - Transition steps in the transition of setpoints: 5
  - Regression control execution cycle: 30 seconds

### A. Scenarios considered for regression control

The performance of the regression control is compared with the conventional $Q(V)$ control since the latter has similar characteristics but does not undergo optimisation. The central OPF and the completely uncontrolled case also serve as a basis for comparison. In this particular test, the dead-band between 0.99 p. u. and 1.01 p. u. was not considered. The simulated scenarios are as follows:
- Case 1: No control, fixed $\cos(\varphi)= 1$ for all RES
- Case 2: Central OPF
- Case 3: Central OPF (normal conditions) + Regression control (fall-back during communication failure)
- Case 4: Central OPF (normal conditions) + Local $Q(V)$ control (fallback solution during communication failure) with a linear slope from 0.95 p. u. to 1.05 p. u.

The performance of the various use cases is explained in Fig. 7.

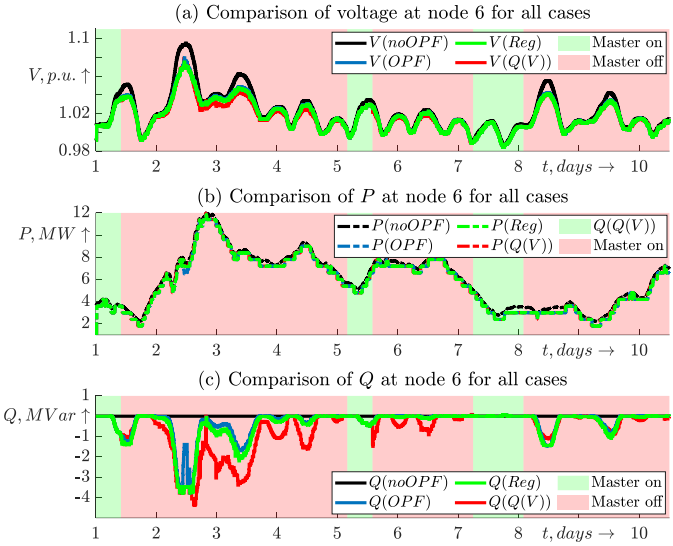

Fig. 7: $P, Q$ infeed of RES and voltage at node 6 with case 3

In general, the central OPF and the regression control should be similar and the transition from central OPF to the local regression control should occur without large jumps in the $P, Q$ setpoints. In comparison, the $Q(V)$ control or no control deviates from the behaviour of the central OPF. Fig. 7 (a) shows the voltage at node 6 at which the DER is connected and the various $P, Q$ setpoints of the DER are represented in Fig. 7 (b) and (c). It can be seen that there are no significant jumps in the subsequent transition from the central OPF to the regression control. In contrast, during the $Q(V)$ control, there are significant changes in the $P, Q$ setpoints of the DER. Accordingly, the DER control curves differ significantly from those in case 2.

At the instant of communication failures, it can be seen that the $Q$ setpoint jumps significantly. There are places where the regression control does not precisely provide exact results similar to central OPF, e.g., at the extremes of the $Q$ curve. In this implementation, SE is configured to take the average measurements over 10 seconds (user-defined). The control is done every 30 seconds and momentary changes in the system voltages are not considered. TABLE 2 shows the mean value of the cost function for each scenario over the entire simulation period.

TABLE 2: Comparison of the mean costs between the cases

| Simulated case | Mean Cost | Diff. to case 1 | Diff. to case 2 |
|---|---|---|---|
| 1: No OPF, $\cos(\varphi)$= 1 | 8.28 | - | 91.57% |
| 2: Normal Operation, central OPF | 4.32 | -47.8% | - |
| 3: OPF in normal case, regression control during failure | 4.39 | -46.9% | 0.9% |
| 4: OPF in normal case, conventional $Q(V)$ control during failure | 4.78 | -42.2% | 10.73% |

It can be seen that switching from operating with a fixed power factor to a central OPF reduces the mean value of the costs by about 47.8 %. When communication failures occur and regression control takes over the determination of the setpoints (case 3), the mean value of the costs increases by only 0.9 % compared to the central OPF. Compared to the uncontrolled scenario in case 1, the mean value of the costs for regression control still decreases by more than 46.9%. Thus, regression control allows the grid to operate in approximately near-optimal conditions.

The use of conventional $Q(V)$ control as a fallback solution in case 4 worsens the mean value of the costs with an increase of more than 10.73 % compared to normal operation (case 2). The mean costs in case 4 are very high compared to central OPF but not higher than in the completely uncontrolled operation (case 1). This also shows that using a conventional $Q(V)$ control as a fallback solution comes with a sudden change in setpoints from central OPF to the local control. According to the results, if there is a very high difference in the $P, Q$ setpoints and the communication fluctuates often, it could challenge the grid stability. The regression control does not have significan jumps in the $P, Q$ setpoints compared to OPF control. An untrained scenario was simulated and it can be seen that neither $Q(V)$ control or the regression control matches the performance of the centralised OPF strategy. Since all the possible cases cannot be trained initially, the results mandate the necessity of training the regression models again for previously untrained scenarios.

### B. Investigation of Offline learning

The results of the regression model retraining are discussed here. In principle, regular retraining can be performed on the field IEDs by periodically executing the regression algorithms. Still, this dynamic process is not the goal of the following investigations. Instead, the results and the effects on the regression models are considered, which results in using such a procedure under the influence of changing system behaviour. The concept for testing the impact of offline learning and retraining of the regression models is shown in Fig. 8.

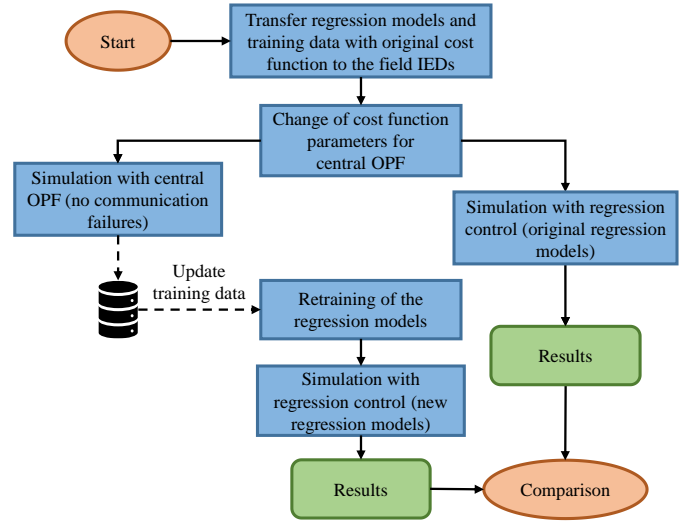

Fig. 8: Test concept for updating the regression models

The regression models of the nearest neighbour regression are used to initialise the field IEDs. The parameters of the OPF cost function are changed on the server to achieve a change in system behaviour. In this case, the OPF was set to use more $P$ than before and there were also some changes to the maximum DER generation capacity. Likewise, changed grid topology or the addition or removal of a load or DER would also be suitable to cause an adaptation of the OPF setpoints. The new values of the cost parameters are given in TABLE 3:

TABLE 3: Change of the cost function parameters

| Parameter | Original Value | New Value |
| --- | --- | --- |
| $c_v$ | 2e³ | 2e³ |
| $c_p$ | 1e6 | 1e5 |
| $c_q$ | 1e4 | 5e4 |

Fig. 9 shows the change of the regression models after the retraining at the operating points $P_{\max} = 60\,\%$ and $P_{\max} = 100\,\%$ for the RES at node 6. It can be compared with Fig. 2 before retraining. It becomes clear that due to the small number of new setpoints, the training data is still mainly characterised by the initialised data points. The regression models adapt only locally and are strongly dependent on the local conditions of the new cost function. The influence of the updated regression models on the system behaviour is investigated in the context of the regression control. The regression control is executed once with the initial and updated regression models. In this way, the performance of the original and retrained regression models can be compared.

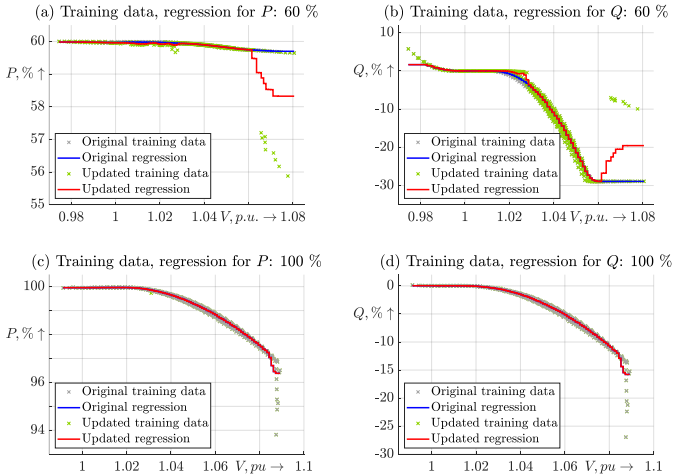

Fig. 9: 20-nearest neighbour original and retrained regression models with training data for 60% $P_{max}$ and 100% $P_{max}$ setpoints.

TABLE 4 compares the costs between the centralised OPF, the original and updated regression models. It can be seen that the mean value of the costs increases by 16.30% when using the original regression models compared to normal operation. On the other hand, in the updated regression models used for the regression control, the mean value increases by only 5.48%. Provided that the periodic offline learning is to be performed on the field IEDs, the duration of the retraining process is of interest. The nearest-neighbour regression takes about 8 seconds to compute all 42 regression models (21 for $P$ and 21 for $Q$).

TABLE 4: Comparison of the mean costs between the cases

| Simulated case | Mean Cost | Diff. to case A |
| --- | --- | --- |
| A: Normal Operation, central OPF | 1.442 | - |
| B: OPF in normal case, original regression models during failure (without retraining) | 1.677 | 16.30% |
| C: OPF in normal case, New regression models during failure (after retraining) | 1.521 | 5.48% |

## VI. CONCLUSION AND OUTLOOK

In this paper, an approach for handling communication failures is developed. For this purpose, a local control using regression models is implemented on the field IEDs as a fallback strategy. The database required for training the regression models is created using the OPF available in MATPOWER, which is run for various operating situations. In addition to performing the regression control, the field IEDs can also regularly adapt the regression models to an updated training database as part of a periodic offline learning process. The performance of the regression control was compared with that of the conventional $Q(V)$ control and found out that the regression control was much better during communication failures in this setup. Thus, the paper suggests using a regression model as a fallback solution during communication failures.

In further tests, additional simulation scenarios can be used to verify the results presented. In particular, the cost function parameters and their influence on the system behaviour are of interest. The parameters used in this implementation have no connection to actual values. Since the cost function was used to evaluate the control, the results can only be applied for this use case and should be adapted for other systems correspondingly.

The dependence of the setpoints on the available generation is considered by creating regression models for the different operating points. In addition, the regression models are created using only the nodal voltage as a predictor and have only one system response with $P$ or $Q$ setpoints. It may be helpful to adjust the regression algorithms to use both the current operating point and the nodal voltage as variables. Also, concerning the system response, a transition can be made to a multidimensional regression model that determines the $P, Q$ setpoints. If dynamic topology changes in the grid during the communication failure, the regression algorithm would not be most efficient.

To counter this, the existing algorithm could be extended to consider possible topology changes by recognising the actual grid topology. The machine learning algorithms did not focus too much on developing the perfect control algorithm. In addition, some assumptions were made to simplify the implementation on the hardware device. While the presented approach showed promising results, the impact due to the regression control was too insignificant to impact the system voltages significantly. This is because only two DERs are assumed to be controllable in the considered benchmark grid. When many DERs are present in the grid and operated during a communication failure, the proposed solution is expected to offer substantial advantages to the bulk power system as it is scalable. This approach of intelligently controlling the DERs will be highly efficient since the coordinated operation of DERs in the distribution grid will help the bulk power system operation enormously.

ACKNOWLEDGMENT

This work was initiated during the research visit of the author to the Power System Laboratory, ETH Zürich in early 2020. The authors gratefully acknowledge the contributions of Prof. Gabriela Hug for her inputs and valuable comments.